\documentstyle[aps,multicol]{revtex}
\begin{document}
\draft
\title{Hyperfine structure in hydrogen and helium ion}
\author{Savely G. Karshenboim$^{a,b}$\cite{email}
and Vladimir G. Ivanov$^{c,a}$}
\address{$^a$ D. I. Mendeleev Institute for Metrology, 198005 St. Petersburg, Russia\\
$^b$ Max-Planck-Institut f\"ur Quantenoptik, 85748 Garching, Germany\\
$^c$ Pulkovo Observatory, 196140, St. Petersburg, Russia}
\date{\today}

\maketitle
\begin{abstract}
QED theory of the hyperfine splitting of the $1s$ and $2s$ state
in hydrogen isotopes and helium-3 ion is considered. We develop an
accurate theory of a specific difference $8E_{\rm HFS}(2s)-E_{\rm
HFS}(1s)$. We take into account fourth order corrections and
nuclear structure effects. The theoretical prediction is now of a
higher accuracy than the experiment is. The study of the
difference provides the most accurate test (on a level of a part
in $10^{8}$) of the QED theory of $1s$ HFS up to date. The theory
agrees with most of the experimental data.
\end{abstract}

\begin{multicols}{2}
\narrowtext

The hyperfine structure (HFS) intervals of the ground state in a
number of neutral atoms and singly charged ions can be measured
with a high accuracy. However, theory even in the case of the
simplest of them (such as hydrogen isotopes and the helium-3 ion)
is essentially affected by nuclear structure effects which
contribute from 30 to 200 ppm and cannot be calculated accurately.
In contrast the $1s$ HFS interval in muonium is calculated with an
uncertainty of about 0.1 ppm and can be used to accurately test
the bound state Quantum Electrodynamics (QED). However, the muonium
calculations involves precision values of the fundamental contants
($\alpha$ and $m_\mu/m_e$) and it would be important to test the
QED calculations for the HFS without interferring with such a
problem.

Study of a specific difference
\begin{equation}\label{difference}
D_{21}=8E_{\rm HFS}(2s)-E_{\rm HFS}(1s)
\end{equation}
provides us with an opportunity to make a test of the QED theory
on a level of accuracy essentially better than 1 ppm
\cite{preliminary}. Such a high accuracy is possible because of an
essential cancelation of nuclear contributions. We report here on
new results for the difference in Eq.~(\ref{difference}). We complete
calculations of the fourth order corrections and find nuclear
structure contributions which remain after cancelation of the
leading effects. Some of the corrections obtained here are bigger
than the experimental uncertainty  \cite{prior} and must be taken
into account.

Our results are found to be in a fair agreement with most
experimental data on hydrogen, deuterium and helium-3. We present
a significant improvement of the theory for $D_{21}$ and
demonstrate that the comparison of theory and experiment
\cite{prior,he1s} for helium tests presently the QED theory of
$1s$ and $2s$ HFS on the highest level, namely one part in $10^8$.
That superceeds the muonium HFS by an order of magnitude.

The hyperfine splitting of an $ns$ state in a hydrogen-like atom with a nuclear mass $M$ and a nuclear spin $I$
can be presented in the form
\begin{equation}
E_{\rm HFS}^{\rm QED}(ns) =
{E_F\over n^3}\cdot\bigl(1+Q_{QED}(ns)\bigr)\;,
\end{equation}
\begin{equation}\label{efermi}
E_F/h =
{8 \over 3} \,Z^3 \alpha^2 \,c\cdot Ry \, {\mu \over \mu_B} \,{ 2 I + 1 \over 2 I}\,
\left({ M\over m+M}\right)^3\;.
\end{equation}
Here $Ry$ is the Rydberg constant, $c$ is the speed of light, $h$
is the Planck constant, $\mu_B$ is Bohr's magneton and $m$ is the
electron mass. The nuclear magnetic moment $\mu$ in our notation
can be negative (if its direction is opposite to the nuclear spin)
and the Fermi energy $E_F$, related to an energy splitting between
the atomic state with total angular moment $F=I+1/2$ and $I-1/2$,
can be negative as well. The QED correction for the HFS interval
in the ground state is (see Ref.~\cite{EGS} for references)
\begin{eqnarray}\label{QED1s}
Q_{QED}(1s)&=&a_e+\left\{\frac{3}{2}(Z\alpha)^2+ \alpha(Z\alpha)\left(\ln2-\frac{5}{2}\right)\right.\nonumber\\
&+&{\alpha (Z \alpha )^2\over \pi}\left[-\frac{2}{3}\ln{1\over(Z\alpha)^2}
\left(\ln{\frac{1}{(Z\alpha)^2}}\right.\right.\nonumber \\
&+&\left.4\ln2-\frac{281}{240}\right) +17.122\,339\ldots \nonumber \\
&-&\left.\left.\frac{8}{15}\ln{2}+\frac{34}{225}\right]+0.7718(4)\,\frac{\alpha^2(Z\alpha)}{\pi}\right\}\,,
\end{eqnarray}
where $a_e$ is the electrons anomalous magnetic moment. A
comparison of the QED calculations with experimental values is
summarized in Table~\ref{T1sHFS}. To compute theoretical values
we use fundamental and auxiliary constants from Refs.~\cite{codata,firestone}. 
The QED expression above does not take
into account any recoil effects. Recoil contributions involve high
momentum transfer \cite{arno} and are essentially affected by the
nuclear structure. In Table~\ref{T1sHFS} we also present data for
the $2s$ state, the theoretical expression for which is similar to
Eq.~(\ref{QED1s}) but some coefficients are different (see below).

One can see that the $1s$ hyperfine structure has been measured
very accurately but any test of the QED calculations is limited by
an essential contribution related to the nuclear structure which
cannot be calculated precisely. In fact the uncertainty of the
nuclear-structure contribution is at least 20\% in hydrogen
\cite{kars99}, and for deuterium the accuracy is not better
\cite{khri}. In the case of tritium and helium-3 ion no results on
the contribution of the nuclear effects has been obtained to the
best of our knowledge. Thus, the pure QED theory is incomplete
because of lack of the nuclear-structure contributions and a
comparison of the QED theory with the experiment in
Table~\ref{T1sHFS} demonstrates how much it is incomplete. 
Our final target is a comparison of $1s$ and $2s$ HFS
intervals and for this reason we do not try to correct the QED
theory for the nuclear effects. Contrary, we compare the pure QED
calculation and experimental data to ``measure'' the nuclear
contribution.

For comparison we presented in Table~\ref{T1sHFS} a theoretical
result on $1s$ muonium HFS \cite{CEK}, which contains the recoil
contributions and even small non-QED terms. Muonium, being a pure
leptonic atom, is free of the nuclear-structure problem, however,
the accuracy of any theoretical calculation is limited to
$10^{-7}$ by the uncertainty of experimental values for
parameters needed to calculate the Fermi energy in Eq.~(\ref{efermi}).
Those are the muon magnetic moment and the fine structure
constant. Below we demonstrate that combining data for the $1s$
and $2s$ hyperfine structure in hydrogen and $^3{\rm He}^+$ we can
go far beyond 1 ppm level \cite{preliminary} and hence develop a
precision test of the QED theory for HFS compatible with the one
related to muonium HFS.

The theory for the specific difference $D_{21}$ in Eq.~(\ref{difference})
up to the third order in units of the Fermi energy was developed some time ago \cite{zwanziger,sternheim,mohr}
\begin{eqnarray}\label{d21qed2}
D_{21}^{(3)}({\rm QED}) &=& (Z\alpha)^2 \,E_F
\times \Biggl\{ {5 \over 8} + {\alpha\over \pi}\,\left[{8 \over 15} \,\ln2 - {7 \over 10}\right]
\nonumber\\
&+& {\alpha\over \pi} \, \left[
\left( {16 \over 3} \,\ln2-7\right)\, \ln(Z\alpha) -5.221\,23\dots\right]
\nonumber\\
&+& {m\over M}\,\biggl[
 -{9 \over 8} +  \left({ \ln2 \over 2}-{7 \over 32}\right)\left(1-
{\mu_B \over \mu} \,{Z\,I\,m \over M}
 \right)\nonumber\\
&-& \left({145 \over 128} - { 7 \over 8} \,\ln2 \right)
{\mu \over \mu_B} \,{M \over m}\,{1 \over {Z\,I}}
\biggr]\Bigg\}\;.
\end{eqnarray}

The nuclear-structure corrections essentially shift the HFS value from its QED prediction.
Three major nuclear effects contribute to the difference
\begin{equation}\label{comparison}
\Delta E_{\rm HFS}^{\rm Nucl}(1s) = E_{\rm HFS}^{\rm exp}(1s)-E_{\rm HFS}^{\rm QED}(1s)
\end{equation}
in Table~\ref{T1sHFS}. Namely they are:
\begin{itemize}
\item the nuclear charge and magnetic moment distribution (that is the biggest effect in the case of hydrogen);
\item a nuclear polarizability contribution (that is the biggest effect in the case of deuterium);
\item nuclear recoil contributions of order $(Z\alpha)(m/M)E_F$ and higher.
\end{itemize}
There is also a correction to the Lamb shift caused by the nuclear structure
\begin{equation}\label{Lamb}
\Delta E_{\rm Lamb}^{\rm Nucl}(1s) = \frac{2}{3}(Z\alpha)^4m^3R_E^2\;,
\end{equation}
where $R_E$ is the nuclear electric charge radius and relativistic
units in which $\hbar=c=1$ are used. When the contributions to HFS
(\ref{comparison}) and the Lamb shift (\ref{Lamb}) are determined, 
one can try to obtain a correction for
difference $D_{21}$. That is possible because most of the
nuclear-structure corrections do not depend on the details of the
atomic structure. Both, the leading contributions to the HFS and
the Lamb shift are of a special factorized form
\begin{equation}\label{NuclPsi}
\Delta E({\rm Nucl}) = A({\rm Nucl})\times \big\vert\Psi_{nl}({\bf r}=0)\big\vert^2\;.
\end{equation}
The energy shift is the product of the nuclear-structure parameter
$A({\rm Nucl})$ and the value of the wave function 
\begin{equation}\label{psi0}
\big\vert\Psi_{nl}({\bf r}=0)\big\vert^2 = \frac{(Z\alpha)^3 m^3}{\pi n^3}\delta_{l0}\;.
\end{equation}
The leading correction to the difference in Eq.~(\ref{difference}) must therefore vanish. The non-vanishing contributions
can be expressed in terms of some effective $\delta$-like potentials
\begin{equation}
V({\rm Nucl})=A({\rm Nucl})\cdot \delta({\bf r})\;.
\end{equation}
The coefficient $A({\rm Nucl})$ can be for various
nuclear contributions calculated (see e.g. Eq.~(\ref{Lamb})) or determined from a
comparison of experiment and a pure QED theory (see e.g.
Eq.~(\ref{comparison})). The result is of the form (cf.
Refs.~\cite{kars99,preliminary,kars})
\begin{eqnarray}\label{d21nuc}
D_{21}({\rm Nucl})&=&\left(\ln2+{3\over16}\right)\cdot(Z\alpha)^2\cdot \Delta E^{\rm Nucl}_{\rm HFS}(1s)\nonumber\\
&+&\left({21\over8}-2\ln2-{3\over 8}\,\zeta\right)\cdot\frac{\Delta E_{\rm Lamb}^{\rm Nucl}(1s)}{(Z\alpha)^2m}\,E_F\;,
\end{eqnarray}
where $1+\zeta ={R_M^2}/{R_E^2}$ is a ratio of quadratic magnetic
and electric nuclear radii. We obtain the nuclear structure
contribution to the $1s$ HFS interval from comparison
in Eq.~(\ref{comparison}) and conservatively estimate the uncertainty as
10\%. The Lamb shift contribution is taken from Eq.~(\ref{Lamb}).

The fourth order corrections to $D_{21}$ have been intensively
studied for the last three years. The logarithmic corrections in
order $\alpha^2(Z\alpha)^2$ and $\alpha(Z\alpha)^2(m/M)$ were
calculated in Ref.~\cite{preliminary} (cf. \cite{Zh,ZP}), the $(Z
\alpha)^3(m/M)$ terms are found below (cf. \cite{Zh}). The only
term calculated previously is the relativistic term of the order
$(Z\alpha)^4E_F$ \cite{breit}.

Partial results on the $\alpha(Z\alpha)^3$ contributions were
found in Ref.~\cite{preliminary}. They are related to effective
non-relativistic potentials which lead to logarithmic
contributions for the $1s$ state HFS \cite{ZP}. The terms in the
same order should also appear from potentials which contain some
derivatives. A complete result on the self energy contribution was
calculated after a suggestion by us in Ref.~\cite{yero2001}. We
report here the completion of the evaluation of the vacuum
polarization effects. We derive an exact result for HFS of the
$2s$ state and a contribution to $D_{21}$ is found via a
comparison with the previously obtained result for the $1s$ state
\cite{jetp00}.

The fourth order contibutions are finally found to be
\begin{eqnarray}\label{d21qed4}
D_{21}^{(4)}({\rm QED})&=&(Z\alpha)^2\,E_F\times\Bigg\{{\alpha^2\over2\pi^2}\,
\left({16\over3}\,\ln2-7\right)\,\ln(Z\alpha)\nonumber\\
&-&{\alpha\over\pi}{2m\over M}\,\left({16\over3}\,\ln2-7\right)\,\ln(Z\alpha)\nonumber\\
&+&\frac{Z\alpha}{\pi}{m\over M}\left({4\over 3}\ln2-2\right)\ln(Z\alpha)\nonumber\\
&+&\alpha(Z\alpha)\Bigl(C_{SE}+C_{VP}\Bigr)+ {177 \over 128}\,(Z\alpha)^2\Bigg\},
\end{eqnarray}
where
\[
C_{SE}(Z=1)=2.07(25)\;,\qquad C_{SE}(Z=2)=2.01(19)\nonumber
\]
and
\[
C_{VP}= \frac{139}{384} + \frac{13}{24} \, \ln2\simeq 0.74\;.\nonumber
\]
The partial results for the constants $C_{SE}$ and $C_{VP}$ that
are obtained in Ref.~\cite{preliminary} contain some misprints. Being
corrected, the partial results ($C_{SE}\simeq 2.5$ and
$C_{VP}\simeq 0.83$) are found to be close to the complete results
above. That confirms an intuitive assumption that the potentials
with derivatives lead to relatively small contributions. Smallness
of terms with derivatives is important for our estimation of
uncertainties of the nuclear-structure corrections.

Let us discuss the uncertainty of the QED expression. The first
two terms in Eq.~(\ref{d21qed4}) are found in the logarith\-mic
approximation and we estimate the next-to-leading terms by a
half-value of the leading contribution. However, in the case of
the third term in Eq.~(\ref{d21qed4}) the situation is more
complicated. First of all, the $(Z\alpha)(m/M)E_F$ corrections to
the $1s$ HFS contain a nuclear-structure dependence presented by
$\ln(mR_E)$. Since we have not included them into the QED
expression (\ref{QED1s}), they are effectively taken into account
as a part of $\Delta E_{\rm HFS}^{\rm Nucl}$. That means that an
essential part of the $(Z\alpha)^3(m/M)E_F$ contribution into
$D_{21}$ is effectively included into $D_{21}({\rm Nucl})$ via
Eq.~(\ref{d21nuc}). However, there are some contributions with loop
momentum of about one electron mass and below which does not
depend on the nuclear structure. They can be enhanced because of a
relatively big magnetic moment (compared to the Dirac value) and
we estimate the uncertainty of the last term in Eq.~(\ref{d21qed4}) as
$(\mu/\mu_B) (Z\alpha)^3E_F$ (cf. Eq.~(\ref{d21qed2})).

All contributions to the difference $D_{21}$ in hydrogen,
deuterium and helium-3 ion are summarized in Table~\ref{Td21new}.
Parameter $\zeta$ is known very badly, but it is not expected to
be much larger than unity and hopefully the $\zeta$-term is
essentially below the uncertainties related to theory and
experiment and thus may be excluded from further considerations.

An essential improvement of the theory is achieved. In previous
papers related to third-order QED corrections
\cite{zwanziger,sternheim} the uncertainty was not spelled out. We
found here a number of corrections exceeding the experimental
uncertainty. We state that after the examination presented here
the theoretical predictions (Table~\ref{Td21new})
\[
D_{21}({\rm theor}) = D_{21}^{(3)}({\rm QED})+D_{21}^{(4)}({\rm QED})+D_{21}({\rm Nucl})
\]
are more accurate than the experiment. Five accurate measurements
performed on three atomic systems are compared with our
calculation in Table~\ref{TSummary}. Four experimental results are
in fair agreement with our theory, but a recent result for
hydrogen \cite{rothery} shows a $1.8\sigma$ discrepancy. The most
important comparison is related to $^3{\rm He}^+$: the $2s$ HFS
was measured most accurately \cite{prior} and its value is also
the most sensitive to higher order corrections (because of larger
$Z$ and larger nuclear contributions).
Because of a fair agreement of our theory with the helium
experiment we expect that in the case of hydrogen the discrepancy
is related to a problem on the experimental side.

We consider comparison of theory and experiment for the difference
$D_{21}$ as a test of a calculation of a state-dependent part of
corrections to $E_{\rm HFS}^{\rm QED}(ns)$ and hence we present in
the last column in Table~\ref{TSummary} a standard deviation
$\sigma$ with respect to the Fermi energy $E_F$, i.~e. to a value
directly related to the 1s HFS. That comparison demonstrates that
study of $D_{21}$ in helium ion provides a more accurate test of
QED than the study of the muonium HFS ($\sigma/E_F\simeq0.1$~ppm)
and indeed of HFS in hydrogen and other atoms with a structured
nucleus. The uncertainty for $D_{21}$ in hydrogen and deuterium is
determined experimentally, while in the case of helium a value of
$\sigma$ contains an essential contribution from theory as well.

Most of so-called QED tests involve in part some other problems such as
\begin{itemize}
\item verification of nuclear models and calculations of nuclear effects and hadronic contributions;
\item tests of consistency of data for fundamental constants (such as muon magnetic moment)
or effective parameters (such as the proton charge radius) related to completely different experiments.
\end{itemize}
The $D_{21}$ theory is free of all these problems. No constants are really involved: an effective value of
$E_{\rm HFS}^{\rm Nucl}(1s)$ related to the nuclear effects arises from HFS theory.
Its contribution being relatively small is under control as well as other nuclear contributions.

It is important to mention that presently there are three crucial
higher-order QED contributions to hydrogenic energy levels:
radiative recoil of the order $\alpha(Z\alpha)^2(m/M)E_F$, pure
recoil of the order $(Z\alpha)^3(m/M)E_F$ and two-loop effects of
the order $\alpha^2(Z\alpha)^6m$. The difference $D_{21}$ is
sensitive to all of them and a progress in its calculation will
therefore contribute into progress in theory of the hydrogen Lamb
shift, muonium hyperfine structure and positronium energy levels.
The most accurate measurement on this difference is related to an
25-years old experiment on helium ions \cite{prior} and we can
hope that some experimental progress to improve the most precise
test of QED theory for the hyperfine structure is possible.

An early part of this work was done during a short but fruitful stay of SGK at University of Notre Dame
and he is very grateful to Jonathan Sapirstein for his hospitality, stimulating discussions and
participation in the early stage of this project. We would like to thank Thomas Udem for useful discussions.
The work was supported in part by RFBR grant 00-02-16718,
Russian State Program ``Fundamental Metrology'' and DAAD.

\begin{table}
\caption{Hypefine structure in light atoms. Hydrogen
result for $1s$ \protect\cite{cjp2000} is an average value over the most accurate data \protect\cite{h1s}.
The difference between experiment and QED theory is denoted by $\Delta E$. It is related to the nuclear contribution.
}
\label{T1sHFS}
\begin{center}
\begin{tabular}{lccc}
Atom, & $E^{\rm exp}_{\rm HFS}$& $E^{\rm QED}_{\rm HFS}$& $n^3\Delta E/E_F$ \\
state&[kHz]&[kHz]&[ppm]\\
\hline
~H, $1s$ & 1\,420\,405.751\,768(1), \protect\cite{cjp2000}& 1\,420\,452& $-\;33$\\
~D, $1s$ & 327\,384.352\,522(2), \protect\cite{d1s}& 327\,339&138\\
~T, $1s$ & 1\,516\,701.470\,773(8), \protect\cite{mathur}& 1\,516\,760& $-\;38$\\
$^3$He$^+$, $1s$ & $-\; 8\,665\,649.867(10)$, \protect\cite{he1s}& $-\;8\,667\,569$& 222\\
\hline
~H, $2s$ & 177\,556.785(29), \protect\cite{rothery}& 177\,562.7 & $-\;33$\\
~H, $2s$ & 177\,556.860(50), \protect\cite{h2s}& &$-\;32$\\
~D, $2s$ & 40\,924.439(20), \protect\cite{d2s}& 40\,918.81 &137\\
$^3$He$^+$, $2s$ & $-\; 1\,083\,354.981(9)$, \protect\cite{prior}& $- \;1\,083\,594.7$& 221\\
$^3$He$^+$, $2s$ & $-\;1\,083\,354.99(20)$, \protect\cite{he2s}& &221\\
\hline
~Mu, $1s$ & 4\,463\,302.78(5), \protect\cite{MuExp}& 4\,463\,302.91(56)& 0.0(1)\\
\end{tabular}
\end{center}
\end{table}

\begin{table}
\caption{Various contributions to $D_{21}({\rm theor})$. Theoretical predictions
depend on parameter $\zeta ={R_M^2}/{R_E^2}-1$.}
\label{Td21new}
\begin{center}
\begin{tabular}{lccc}
Value         &  H   & D  & $^3$He$^+$ \\
\hline
$D_{21}^{(3)}({\rm QED})$  [kHz] & 48.937 &  11.305\,6 & $-\;1\,189.262$\\
$D_{21}^{(4)}({\rm QED})$  [kHz] & 0.018(3) &  0.004\,3(5) & $-\;1.137(53)$\\
$D_{21}({\rm Nucl})$ [kHz] & $-\;0.002$ & 0.002\,6(2)   & 0.331(36) \\
 &  $-\; 10^{-4}\;\zeta$  & $-\;10^{-4}\; \zeta$ & $+\;0.009\;\zeta$ \\
 \hline
$D_{21}({\rm theor})$ [kHz] & 48.953(3) &  11.312\,5(5) & $-\;1\,190.067(63)$  \\
 &  $-\; 10^{-4}\;\zeta$  & $-\;10^{-4}\;\zeta$ & $+\;0.009\;\zeta$ \\
\end{tabular}
\end{center}
\end{table}

\begin{table}
\caption{Precision tests of QED theory for $D_{21}$. The final standard deviation $\sigma$ includes
contributions from both: theory and experiment. $\Delta D_{21}$ stands for the difference of experiment
and theory. References for
the $D_{21}$ are presented for experiment for the $2s$ HFS. We put here $\zeta=0$.
}
\label{TSummary}
\begin{center}
\begin{tabular}{lcccc}
Atom  & $D_{21}$(exp) & $D_{21}$(theor) &  $\Delta D_{21}/\sigma$ & $\sigma/E_F$ \\
 & [kHz] & [kHz]  &   & [ppm] \\
\hline
~H  &  48.53(23), \protect{\cite{rothery}} & 48.953(3)  & $-\;1.8$ & 0.16 \\
~H  &  49.13(40), \protect{\cite{h2s}} & & 0.4 &  0.28 \\
~D  &  11.16(16), \protect{\cite{d2s}} & 11.312\,5(5) & $-\;1.0$ & 0.49 \\
$^3{\rm He}^+$  & $-\;1\,189.979(71)$, \protect{\cite{prior}} & $-\;1\,190.067(63)$& 0.9& 0.01 \\
$^3{\rm He}^+$  & $-\;1\,190.1(16)$, \protect{\cite{he2s}} &  &  0.0 & 0.18 \\
\end{tabular}
\end{center}
\end{table}

\end{multicols}

\end{document}